\documentstyle[11pt,amssymb,amsfonts,amsthm,amssymb,amscd,amstext,epsfig]{article}

\def\ben{\begin{equation}}
\def\een{\end{equation}}
\def\half{{\textstyle{1\over2}}}
\def\qtr{{\textstyle{1\over4}}}
   \let\d=\delta 
   \let\k=\kappa

\def\nn{\nonumber}

\let\pa=\partial
\def\be{\begin{equation}}
\def\ee{\end{equation}}
\def\ba{\begin{array}}
\def\ea{\end{array}}

\def\dalemb#1#2{{\vbox{\hrule height .#2pt
        \hbox{\vrule width.#2pt height#1pt \kern#1pt
                \vrule width.#2pt}
        \hrule height.#2pt}}}

\newcommand{\bea}{\begin{eqnarray}}
\newcommand{\eea}{\end{eqnarray}}

\def\R{{{\Bbb R}}}

\def\Z{{{\Bbb Z}}}

\begin{document}
\begin{flushright}
NSF-KITP-06-124 \\
hep-th/0612159
\end{flushright}

\begin{center}
\vspace{1cm} { \LARGE {\bf Anyonic strings and membranes in AdS space \\
and dual Aharonov-Bohm effects}}

\vspace{1.1cm}

Sean A. Hartnoll

\vspace{0.8cm}

{\it KITP, University of California Santa Barbara \\
 CA 93106, USA } \\
 {\tt hartnoll@kitp.ucsb.edu} \\

\vspace{2cm}

\end{center}

\begin{abstract}

It is observed that strings in $AdS_5 \times S^5$ and membranes in
$AdS_7 \times S^4$ exhibit long range phase interactions. Two well
separated membranes dragged around one another in $AdS$ acquire
phases of $2 \pi/N$. The same phases are acquired by a well separated F and
D string dragged around one another. The phases are shown to
correspond to both the standard and a novel type of Aharonov-Bohm
effect in the dual field theory.

\end{abstract}

\pagebreak
\setcounter{page}{1}

\section{Introduction}

It is by now well established that point particles in 2+1
dimensions may have fractional, or anyonic, statistics
\cite{Leinaas:1977fm,Wilczek:1981du,Wilczek:1982wy,
Wilczek:1983cy}. These are particles that do not obey either
Bose-Einstein or Fermi-Dirac statistics. Under exchange of two
identical anyons, the wavefunction of the system does not get
multiplied by $\pm 1$, but may rather change by an arbitrary,
fixed, phase. A physical example of anyons are the quasiparticle
or quasihole excitations of a fractional quantum hall fluid
\cite{Halperin:1984fn}.

Anyons may be physically understood as particles with a long range
phase interaction. In 2+1 dimensions a particle may be charged
under an abelian Chern-Simons gauge potential. The Chern-Simons
interaction attaches a magnetic flux to all electric point
charges. Therefore when two such charges are dragged around one
another, at arbitrarily large distances, they acquire a phase
through the Aharonov-Bohm effect.

The possibility of anyonic phases in two spatial dimensions is
directly associated with the topology of the two particle
configuration space. In particular, $H_1(\R^2 \setminus
\{\text{0}\}) = \Z$, so the path of a particle that is dragged
around another is topologically nontrivial. In higher dimensions,
these paths are homologically trivial and hence anyonic phases are
not possible for point particles. However, higher dimensional
objects such as strings and membranes do have nontrivial
configuration spaces in higher dimensions. In particular,
$H_2(\R^4\setminus \R) = H_3 (\R^6 \setminus \R^2) = \Z$, which
implies that strings in 4+1 dimensions and membranes in 6+1
dimensions can describe topologically nontrivial cycles as they
are dragged around one another. Some previous work on the
configuration space of strings in 3+1 dimensions may be found in
\cite{Aneziris:1990gm,Aneziris:1992aj}. The 3+1 dimensional
analogue of the physics to be considered here would involve a
string linking a point source, c.f. \cite{Bowick:1988xh}.

The next observation is that two of the most basic backgrounds of
string and M theory, the $AdS_5 \times S^5$ solution of type IIB
string theory and the $AdS_7 \times S^4$ solution of M theory,
realise the potential for anyonic behaviour of strings and
membranes. The strings/membranes need to be well separated
relative to the size of the sphere in the background, so that the
physics is effectively five/seven dimensional, respectively. The
two ingredients that make nontrivial phases possible are firstly the
nonlinear Chern-Simons like terms in the supergravity actions and
secondly the presence of a background flux in the solution. The
resulting mechanism is essentially identical to that for point charges in 2+1
dimensional Chern-Simons theory.

These $AdS$ backgrounds are of special interest because they are
supergravity duals of conformal field theories
\cite{Maldacena:1997re}. After explicitly exhibiting the
advertised long range phase interactions in these backgrounds
below, we will show that, in the case when one of the strings or
membranes reaches the boundary of $AdS$, then the bulk anyonic
behaviour has a dual interpretation as a nonabelian Aharonov-Bohm
effect. If neither of the objets reach the boundary, we show how
the phase may be dually understood as a novel type of Aharonov-Bohm
effect involving intersecting flux tubes.

\section{Anyonic membranes in $AdS_7 \times S^4$}

Start with the case of a single M2 brane coupled to eleven
dimensional supergravity. The physics we are studying only depends
upon the action for the three form $C$ field. Letting $G = dC$,
the required action is
\be
S_{\text{C field}} = - \frac{1}{4 \k_{11}^2} \int \left( G \wedge
\star G + \frac{1}{3} C \wedge G \wedge G \right) + T_{M2}
\int_{M2} C \,.
\ee
Using $2 \k_{11}^2 = (2\pi)^8 l_{11}^9$ and $T_{M2}^{-1} =
(2\pi)^2 l_{11}^3$ one obtains the equations of motion
\be
d \star G + \frac{1}{2} G \wedge G - (2\pi l_{11})^6 \d(x) = 0 \,.
\ee
Here $l_{11}$ is the eleven dimensional Planck length and $\d(x)$
is the Poincar\'e dual eight form to the worldvolume of the
membrane. Recall that the Poincar\'e dual is defined by
\be
\int_{M2} C = \int C \wedge \d (x) \,,
\ee
for any $C$. The second integral here is over the whole spacetime.

We wish to consider the effect of a single membrane in the $AdS_7
\times S^4$ background. This background has flux
\be
G^{(0)} = 3\pi N l_{11}^3 \text{vol}_{S^4} \,,
\ee
where $\text{vol}_{S^4}$ is the volume form on a unit radius four
sphere. The effect of the membrane is to shift this background
flux by a small amount. Let us set $G \to G^{(0)} + G$ and work to
linearised order in $G$. The equation of motion becomes
\be\label{eq:m2}
d \star G + N (2\pi l_{11})^3 \frac{3}{8 \pi^2} \text{vol}_{S^4}
\wedge G - (2\pi l_{11})^6 \d(x) = 0 \,.
\ee
Place this membrane in the $AdS_7$ part of the geometry.

Now consider a second membrane in the same background. Let us
adiabatically drag this membrane along a closed spatial curve that
is within $AdS_7$ and encircles the first membrane, ending up back
where it started. This is well defined as the first membrane is
occupying two spatial dimensions and is therefore a point in the
transverse four spatial dimensions in $AdS_7$, that can be
encircled by a three volume. Denote by $\pa \Sigma$ the
worldvolume of this second membrane. This process is illustrated in
figure 1.

\begin{figure}[h]
\begin{center}
\epsfig{file=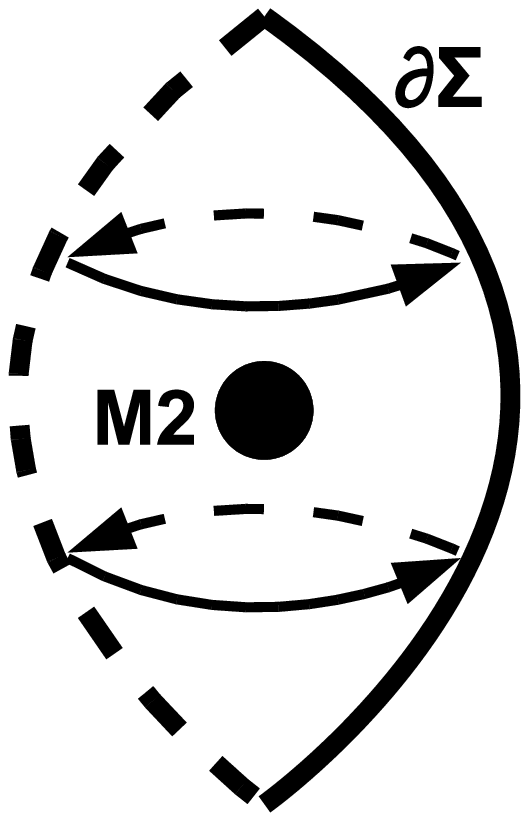,width=4cm}
\end{center}

\noindent {\bf Figure 1:} $M2$ denotes the first membrane and $\pa \Sigma$
is the worldvolume traced out by the second membrane.

\end{figure}

More precisely, a spatial slice of $AdS_7$ has topology $\R^6$.
Removing the space occupied by the first membrane gives topology
$\R^6 \setminus \R^2$. However $H_3 (\R^6 \setminus \R^2) =
\Z$, and it is this nontrivial three cycle we are dragging the
second membrane around.

The phase acquired by the partition function of the second
membrane upon completion of the closed loop is computed as follows
\bea \label{eq:m2shift}
\Delta \Phi_M & = & \frac{2 \pi}{(2 \pi l_{11})^3} \int_{\pa \Sigma}
C\nn \\
 & = & \frac{2 \pi}{(2 \pi l_{11})^3} \frac{3}{8 \pi^2}
\int_{\pa \Sigma \times S^4} C \wedge \text{vol}_{S^4} \qquad\qquad [\text{as} \quad R \to \infty] \nn \\
& = & \frac{2 \pi}{(2 \pi l_{11})^3} \frac{3}{8 \pi^2}
\int_{\Sigma \times S^4} G \wedge \text{vol}_{S^4} \nn \\
& = & - \frac{2\pi}{N (2 \pi l_{11})^6} \int_{\Sigma
\times S^4} \left[ d \star G - (2 \pi l_{11})^6 \d(x) \right] \nn
\\
& = & \frac{2\pi}{N} - \frac{2 \pi}{N (2 \pi l_{11})^6}
\int_{\pa \Sigma \times S^4} \star\, G \,.
\eea
In the second step we have required that the minimum spatial
distance between the two membranes, $R$, be large. At large
distances, the $C$ field in $AdS_7$ sourced by the first membrane
is independent of the coordinates on the $S^4$. This fact permits
us to wedge $C$ with $\text{vol}_{S^4}$. The higher harmonics of
the $C$ field decay exponentially at spatial distances beyond the
size of the $S^4$, which is the $AdS$ scale, due to the Laplacian
term $d \star d C$ in (\ref{eq:m2}). In the appendix we have
exhibited this exponential falloff explicitly by solving a toy
model for (\ref{eq:m2}): a point charge in $\R^{1,2}\times S^1$
with analogous Chern-Simons like couplings. This intuition is not
confined to flat space. The higher harmonics acquire masses in
$AdS_7$, and massive fields on $AdS$ decay exponentially in
geodesic distance, see e.g. \cite{Allen:1985wd,Naqvi:1999va,
Bena:1999be}.

In order to evaluate the expression for the phase
(\ref{eq:m2shift}), we should solve the equation of motion
(\ref{eq:m2}) for $G$. However, we can immediately note that at
long distances from the source membrane, the topological mass term
in (\ref{eq:m2}) is important and the field strength dies off
exponentially. Hence the flux term in the last line of
(\ref{eq:m2shift}) is negligible. This falloff in $G$ occurs even
for the homogeneous mode on $S^4$ (the homogeneous $C$ that
survives at long distance has $G=dC=0$). Thus we obtain the phase
\be\label{eq:m2limits}
\Delta \Phi_M = \frac{2 \pi}{N} \qquad \qquad [\text{as}\quad R \to \infty] \,.
\ee
Rephrasing this result in terms of membrane exchange, in which the
second membrane is only taken `half way' around the first, and
adding together the phases acquired by the two membranes, we have
found that well separated membranes in $AdS_7 \times S^4$ obey
anyon statistics with angle $2\pi/N$.

At short distances the full eleven dimensional geometry must be
considered and there is no effective linking of membranes.
Furthermore, the derivative term in (\ref{eq:m2}) will be
important. It may seem surprising that the phase depends on the
distance between the membranes. However, the same behaviour occurs
in Maxwell-Chern-Simons theory in three dimensions
\cite{Shizuya:1990ju}. The effect of the Maxwell term is to give
the magnetic flux tubes attached to the anyons a finite width.
When the charges are close enough together for these to overlap,
the fractional statistics is lost. The characteristic feature
of anyons in any case is the long (infinite) range phase interaction,
independently of short distance physics.

To gain intuition for this system, it is useful to solve the
equation for a topologically massive four form field strength on
$\R^{1,6}$. This is done in the appendix. The exponential decay of
the flux at large distances may be seen very explicitly. The
propagator for a topologically massive form in $AdS$ has been
computed in \cite{Bena:2000fp} and exhibits exponential decay with
geodesic distance.

\section{Long range phase interaction for strings in $AdS_5 \times S^5$}

Consider a single D or F string coupled to type IIB supergravity.
We are interested in an effect due to the two form potentials in
the theory. The action for these forms, in Einstein frame, is
\bea
S_{B_2, C_2} &= & - \frac{1}{4 \k^2_{10}} \int \left(e^{-\Phi} H_3
\wedge \star H_3 + e^{\Phi} \tilde F_3 \wedge \star \tilde F_3 +
\frac{1}{2} \tilde F_5 \wedge \star \tilde F_5 + C_4
\wedge H_3 \wedge F_3 \right) \nonumber \\
& - & T_{F1} \int_{F1} B_2 - \mu_{1} \int_{D1} \left( C_2 - C B_2
\right) \,,
\eea
where $H_3 = d B_2$, $F_3 = d C_2$, $\tilde F_3 = F_3 - C H_3$ and
$\tilde F_5 = F_5 - C_2 \wedge H_3$. The self duality condition
$\tilde F_5 = \star \tilde F_5$ is imposed on the equations of
motion. We have included contributions from both an F and D string
to the action. To start with, we are interested in one or the
other.

The equations of motion for the two form potentials $C_2$ and
$B_2$ are found to be
\be
d \star \left( e^{\Phi} \tilde F_3 \right) - F_5 \wedge H_3 - 2
\k^2_{10}\mu_{1} \d (x_{D1}) = 0 \,,
\ee
and
\be
d \star \left( e^{-\Phi} H_3 - C e^{\Phi} \tilde F_3 \right) + F_5
\wedge F_3 -  2 \k^2_{10}T_{F1} \d (x_{F1}) +
2 \k^2_{10}\mu_{1} C \d (x_{D1}) = 0 \,.
\ee
In these expressions, $\d (x)$ denotes the Poincar\'e dual eight
form to the string worldvolume.

We want to consider the effect of a single D or F string in the
$AdS_5 \times S^5$ background. This background has five form flux
\be
F^{(0)}_5 = 16 \pi N l_s^4 \left( \text{vol}_{S^5} + \star
\text{vol}_{S^5} \right) \,,
\ee
and furthermore has constant dilaton and axion: $e^{\Phi} = g$ and
$C = \theta/2\pi$. Here $\text{vol}_{S^5}$ is the volume form on
an $S^5$ of unit radius and $l_s$ is the string length. The string
source introduces a small amount of three form flux which we
describe by linearising the equations of motion above. Using
$T_{F1} = \mu_{1} = 1/(2\pi l_s^2)$ and $2 \k^2_{10} = (2\pi)^7
l_s^8$, and writing the equations in a more symmetric way, gives
\be\label{eq:s1}
g d\star \tilde F_3 - \frac{N (2\pi l_s)^4}{\pi^3}
\left(1 + \star \right) \text{vol}_{S^5} \wedge H_3 -
(2\pi l_s)^6 \d (x_{D1}) = 0 \,,
\ee
and
\be\label{eq:s2}
\frac{1}{g} d\star H_3 + \frac{N (2\pi
l_s)^4}{\pi^3} \left( 1 + \star \right) \text{vol}_{S^5} \wedge
\tilde F_3 - (2\pi l_s)^6 \d (x_{F1}) = 0
\,.
\ee
Note that here
\be
\tilde F_3 = F_3 - \frac{\theta}{2\pi} H_3 \,.
\ee

From the structure of the equations of motion (\ref{eq:s1}) and
(\ref{eq:s2}), we see that we can obtain nontrivial phases for
strings in $AdS_5$ in the same way as we did for membranes in
$AdS_7$. This will occur if we drag an F string around a D string,
or vice versa. However, there is no phase associated
with dragging an F string about another F
string or a D string about another D string. For two D strings for
instance, we can compute the phase as follows
\bea\label{eq:dshift}
\Delta \Phi_D & = & - \frac{2\pi}{(2\pi l_s)^2} \int_{\pa \Sigma}
\left( C_2 - \frac{\theta}{2\pi} B_2 \right) \nonumber \\
 & = & - \frac{2\pi}{(2\pi l_s)^2} \frac{1}{\pi^3} \int_{\Sigma \times S^5}
 \tilde F_3 \wedge \text{vol}_{S^5} \qquad\qquad [\text{as} \quad R \to \infty] \nonumber \\
 & = & - \frac{2 \pi}{N g (2\pi l_s)^6} \int_{\pa \Sigma \times S^5} \star H_3 \,.
\eea
We have used the same notation as previously for membranes. Now
$\pa \Sigma$ is a two cycle surrounding the first D string in
$AdS_5$. This is well defined as $H_2(\R^4\setminus \R) = \Z$.
Unlike the previous case of membranes, this flux does not decay
exponentially at large distances but is instead identically zero.
This can be seen by considering the electric and magnetic parts
$H_3 = E \wedge \sqrt{-g_{tt}} dt + B$ and $\tilde F_3 = \tilde E
\wedge \sqrt{-g_{tt}} dt + \tilde B$, where $t$ is a static time coordinate for $AdS$.
From the equations of motion we have that
\bea\label{eq:nosource}
\frac{1}{g} d \star_9 E
 - \frac{N (2\pi l_s)^4}{\pi^3} (1+\star) \text{vol}_{S^5} \wedge \tilde B & = & 0
 \,,\nonumber \\
g d \star_9 \sqrt{-g_{tt}} \tilde B
 + \frac{N (2\pi l_s)^4}{\pi^3} (1+\star) \text{vol}_{S^5} \wedge \sqrt{-g_{tt}} E & = &
 0\,.
\eea
There are no source terms in these equations, which are therefore
solved by $E = \tilde B = 0$. It follows from (\ref{eq:dshift})
that $\Delta \Phi_D = 0$.

As for the membranes, more complete intuition for the coupled equations
of motion (\ref{eq:s1}) and (\ref{eq:s2}) may be gained from solving
the equations for similarly coupled topologically massive three form
field strengths on $\R^{1,4}$. This is done in the appendix.

Considering $(p,q)$ strings does not alter the situation. The
relative minus sign between the topological mass terms in
(\ref{eq:s1}) and (\ref{eq:s2}) means that the effects of the F
and D charges cancel and there is no long range phase interaction.
Alternatively, this follows from $SL(2,\Z)$ duality of the D
string results. Therefore, the various strings in $AdS_5$ do not
have anyon statistics. However, there is an infinite range phase
interaction between F and D strings. In this sense we might call
these strings anyonic. The phase shift in the partition function
of a D string dragged around an F string is
\bea
\Delta \Phi_{F-D} & = & - \frac{2 \pi}{(2\pi l_s)^2}
\frac{1}{\pi^3} \int_{\Sigma \times S^5} \tilde F_3 \wedge \text{vol}_{S^5}
\qquad\qquad [\text{as} \quad R \to \infty] \nonumber \\
 & = & \frac{2 \pi}{N} - \frac{2 \pi}{g N (2\pi l_s)^6}
 \int_{\pa \Sigma \times S^5} \star H_3 \,,
\eea
where $\pa \Sigma$ is as usual the worldvolume swept out by the D
string. The electric flux contribution in the previous equation is
determined from (\ref{eq:nosource}) except that now there is a
delta function source term due to the F string in the background.
As for the case of membranes, the topological mass terms imply that
the flux decays exponentially at large distances giving
\be\label{eq:FD}
\Delta \Phi_{F-D} = \frac{2 \pi}{N} \qquad \qquad [\text{as}\quad R \to \infty] \,.
\ee

\section{Dual field theory implications}

The various membranes and strings we have discussed are dual to
specific states (operators) in the dual conformal field theories.
The F and D strings are dual to electric and magnetic flux tubes,
respectively \cite{Balasubramanian:1998de}. The M2 branes are dual
to interesting analogous extended field configurations in the six
dimensional $(2,0)$ theory.

A simple case to map to the dual theory is when one of the strings
or membranes extends to the $AdS$ boundary. In this case the
endpoint of the string or membrane is an external point or string
charge, respectively, in the dual theory. Consider the case of an
F string that extends to the boundary. Dragging this string around
a D string in the bulk is mapped onto an external electric charge
that is dragged around a magnetic flux tube on the boundary. Such
a charge will acquire a phase due to the (nonabelian)
Aharonov-Bohm effect.

In fact, we can recover precisely the phase we found from the bulk
computation. The total flux along the magnetic flux tube is
$2\pi$, which follows from the Dirac quantisation condition.
However, this flux is in the (decoupled) diagonal $U(1)$ of
$U(N)$, so we can write it as $2\pi/N$ times the identity $N
\times N$ matrix. The external electric charge is in some specific
$U(1)$ of $SU(N)$ and therefore is only sensitive to one component
of the flux. Thus from the Aharonov-Bohm effect we obtain the
phase $2\pi/N$, in agreement with the bulk computation.

The Aharonov-Bohm setup just described is illustrated on the left hand side of
figure 2. Note that the string needs to return to the boundary as
it has nowhere to end in the $AdS$ bulk. The figure depicts an F
string that extends to the boundary being dragged around a D
string in the bulk. Note also that the surface linking the D
string is not strictly closed, as it does not close off at the
boundary. However, this finite sized `hole' is infinitely far away
from the source and no flux escapes through it.

\begin{figure}[h]
\begin{center}
\epsfig{file=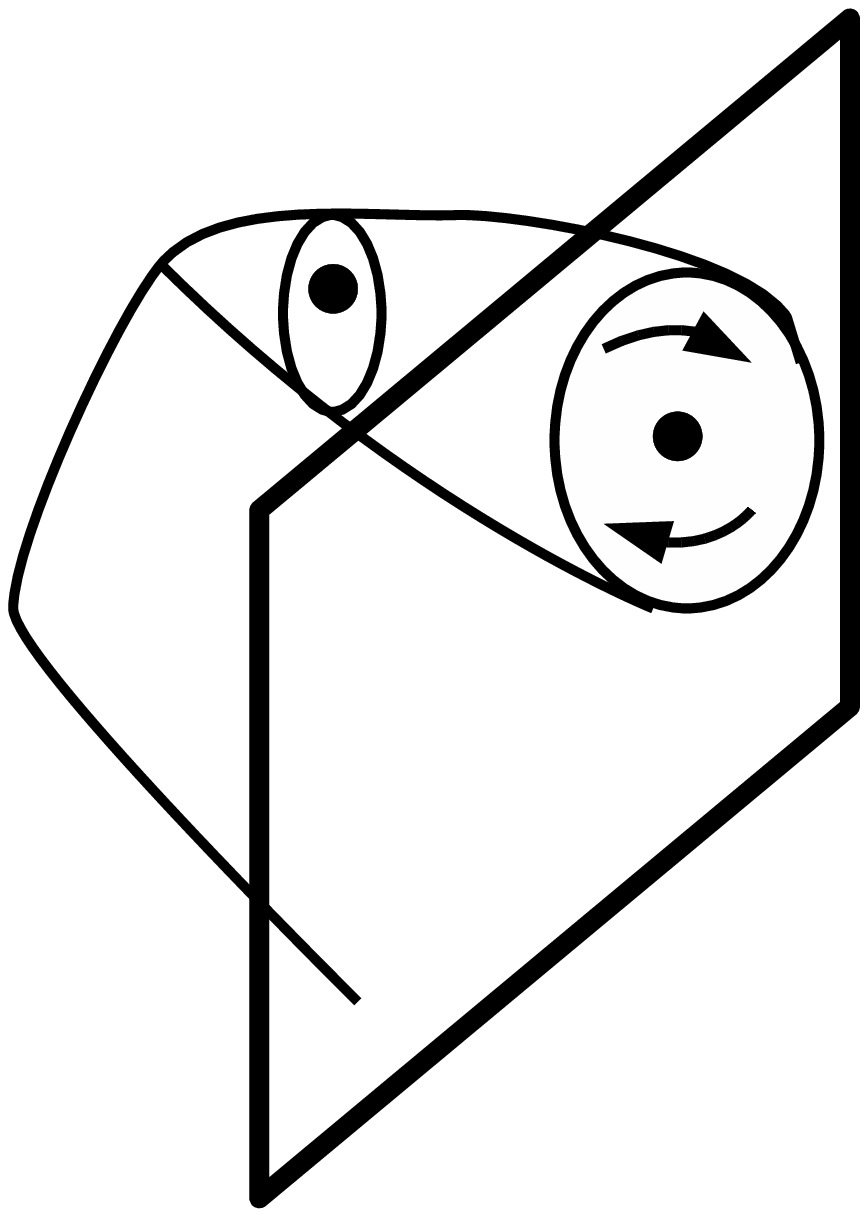,height=6cm}(a) \hspace{2cm}
\epsfig{file=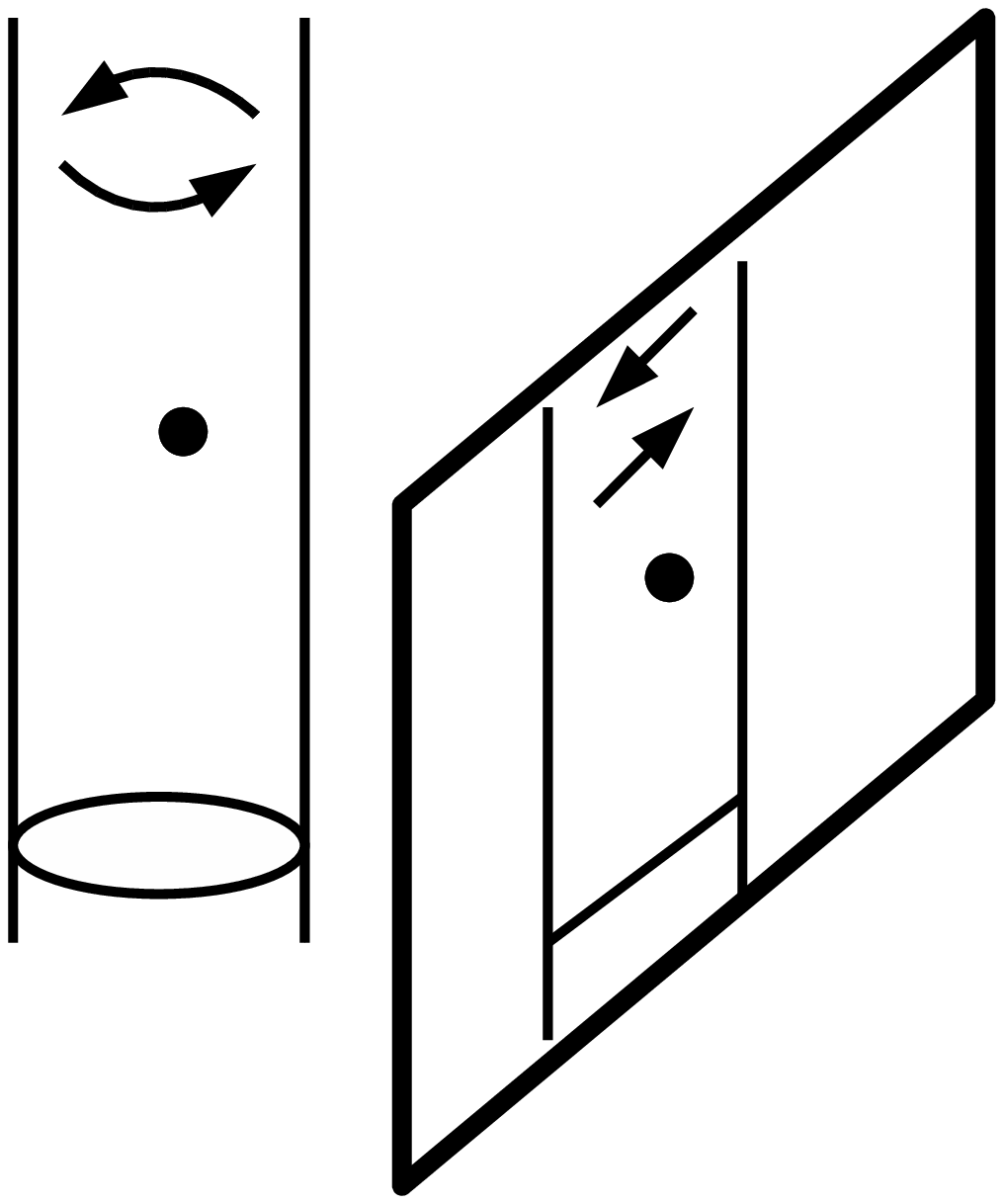,height=6cm}(b)
\end{center}

\noindent {\bf Figure 2:} Schematic representation of $AdS_5$
together with its conformal boundary. Time and one spatial
dimension are suppressed. The dot denotes a D string and its
projection onto the boundary. The lines denote an F string dragged
around the D string. In (a) the F string extends to the boundary
of $AdS$, whereas in (b) it does not.
\end{figure}

The right hand side of figure 2 illustrates the case in which the
F string does not extend to infinity. Projecting the motion onto
the boundary, this implies that the magnetic and electric flux
tubes intersect twice, as one is dragged through the
other. We will now understand why this process picks up a phase.

The first step is to note that an infinitessimally thin electric flux tube dragged
adiabatically around a closed loop, to give a surface $\Sigma$,
acquires a phase $\int_{\Sigma} F$. This can been seen
as a consequence of the fact that adding this term to the action
results in the Maxwell equation
\be
d \star F = d \d(x) \,.
\ee
This is solved by $ \star F = \d(x)$, which precisely describes a flux tube
at $x$ carrying an electric field. Therefore, $\int F$ is the
correct action for an electric flux tube and leads to a phase in the
partition function if the flux tube is dragged along some surface $\Sigma$.
Similarly, the action for a magnetic flux tube is $2\pi \int \star F$.

A test of the proposed phase is to consider a reverse
Aharonov-Bohm effect, in which a magnetic flux tube is dragged
around an electric charge. Then we obtain $2\pi \int_{\Sigma}
\star F = 2\pi$, recovering the same phase as in the standard
Aharonov-Bohm setup, as we should expect.

Given this phase, we can consider the process of interest in
figure 2b. The spatial worldsheet $\Sigma$ of the electric flux
tube intersects the magnetic flux tube twice. These intersections
contribute to the phase $\int_{\Sigma} F$. As the
intersections have opposite orientation, the total magnetic flux
through $\Sigma$ is zero. However, the F string is located at
different radial locations in $AdS$ at the two intersections. This
implies that the cross sectional width of the flux tube is
different at the two points, being thinner when the F string is
nearer the boundary \cite{Balasubramanian:1998de}. In particular,
if $L^{(e)}_{1,2}$ denotes the width of the electric flux tube at
the two intersections and $L^{(m)}$ denotes the constant width of
the magnetic flux tube, then we have
\be
L^{(e)}_1 \ll L^{(m)} \ll L^{(e)}_2 \,,
\ee
corresponding to the fact that the dual strings are separated at radial
distances much larger than the AdS scale.

The formula for the phase we presented assumes a thin flux tube.
For the intersection near the boundary we have $L^{(e)}_1 \ll
L^{(m)}$: the electric flux tube is much thinner than the magnetic
flux passing through it. The formula should be reliable here.
However, at the other intersection, $L^{(m)} \ll L^{(e)}_2 $ and
the electric flux is much more spread out than the magnetic flux.
This suggests that we should swap pictures and view this second
intersection as a thin magnetic flux tube that is being dragged
(in the opposite direction) through electric flux. We need to
carefully keep track of relative orientations as we do this, as
illustrated in figure 3. The magnetic flux tube will then pick up
a phase $\int \star F$, the electromagnetic dual of the phase for the
electric tube. Now recall that under electromagnetic duality ${\bf
E} \to {\bf B}$ but ${\bf B} \to - {\bf E}$. Applying this map to
the new setup in figure 3, we find that the phase acquired is the
same as that from the first intersection. In particular, the sign
is the same. The phases from the two intersections add and do not
cancel.

\begin{figure}[h]
\begin{center}
\epsfig{file=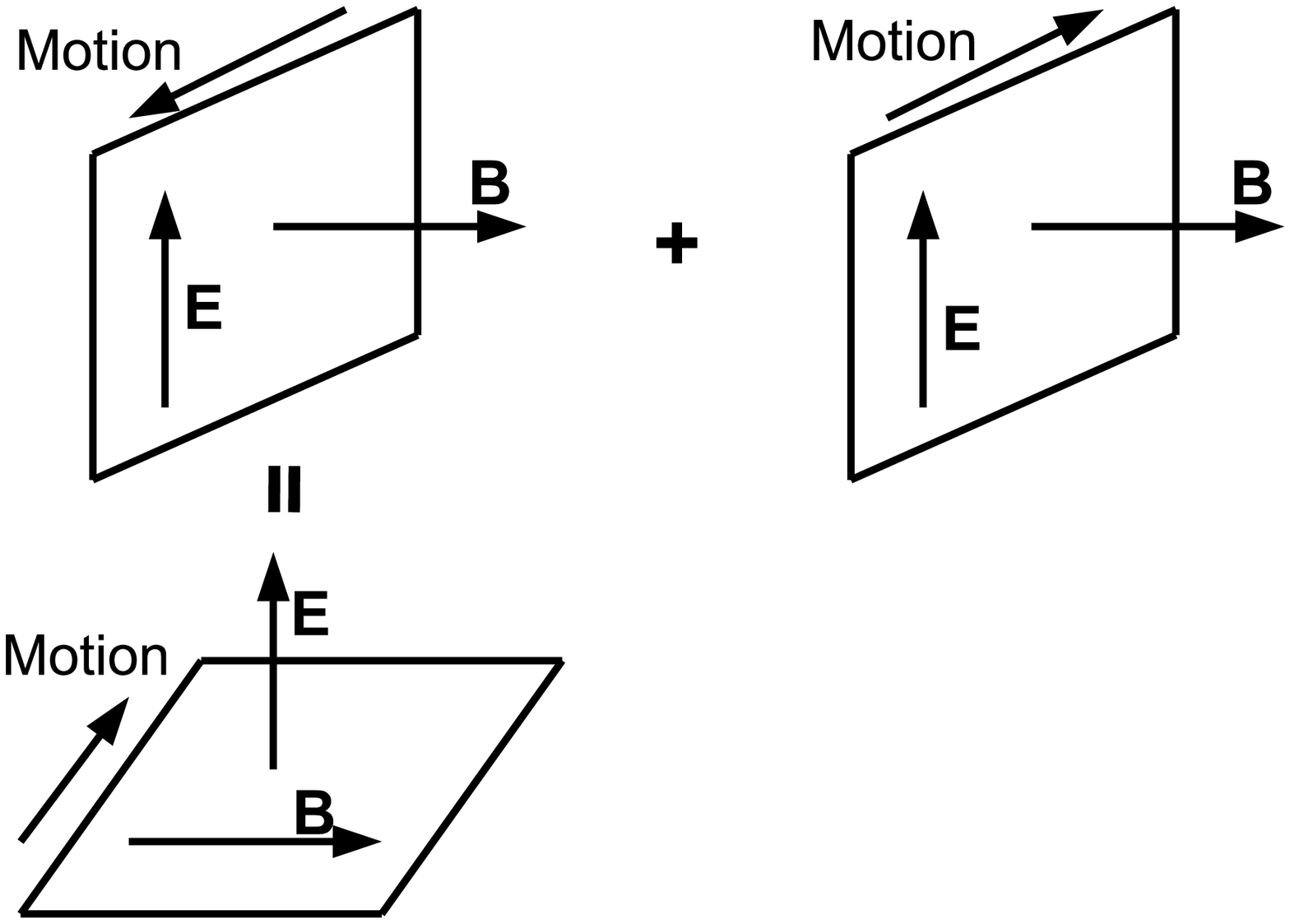,height=7cm}
\end{center}

\noindent {\bf Figure 3:} The top line shows the two intersections of
the electric flux tube with magnetic flux. The second line shows
how one of these should be reinterpreted as a magnetic flux tube
intersected by electric flux.
\end{figure}

We can be more quantitative, and recover the phase found from the
bulk computation. We have seen that the two intersections give the
same phase, so focus on one of them. The phase acquires a $2\pi$
as previously from the relative normalisation of electric and
magnetic flux. A $1/N$ is acquired due to two factors of $1/N$
coming from the distribution of the flux in each flux tube amongst
$N$ diagonal entries of the $U(N)$ matrix, as above, together with
a factor of $N$ coming from summing over the contributions from
each colour. Thus each intersection gives $2\pi/N$ and the total
phase is $4\pi/N$. This is also the total phase obtained in the
bulk, where there is a $2\pi/N$ phase for the partition function
of each string. An assumption underlying this matching is the
following: given that at each intersection one
of the tubes was very spread out and hence thought of as background flux,
the total phase acquired at each intersection is $2\pi/N$, as
opposed to this phase being acquired by each of the tubes
separately (which would lead to a total phase of $8\pi/N$).

The case of membranes works identically. For instance, the string
boundary of a membrane ending on the M theory fivebrane is self
dual under the two form potential of the fivebrane theory. There
is thus only one basic flux tube in the theory, and an external
self dual string source dragged around such a flux tube will
acquire an Aharanov-Bohm phase.

\section*{Acknowledgements}

I would like to thank Albion Lawrence, David Mateos, Rob Myers,
Joe Polchinski and David Tong for helpful and enjoyable
discussions. This research was supported in part by the National
Science Foundation under Grant No. PHY99-07949.

\appendix

\section{Topologically massive forms in flat space}

\subsection{Particle source in $\R^{1,2} \times S^1$}

Write the metric as
\be
ds^2 = -dt^2 + dr^2 + r^2 d\phi^2 + d\theta^2 \,.
\ee
Take $\theta$ to have period $2\pi$. Place the point source at $r
= \theta = 0$. We want to solve the following equation for the one
form potential $A$
\be
d \star F + d\theta \wedge F + 4 \pi^2 \d(x) = 0 \,,
\ee
where $F = dA$. The equation may be straightforwardly solved in
terms of Bessel functions
\bea
A & = & \left(1-K1(r) r \right) d\phi + K_0(r) dt \, \nonumber \\
& + & \sum_{
  m_\pm \geq 1 } C_{m_\pm} e^{i m_{\pm}
\theta} \left[\pm K_1 \left( (\sqrt{m^2_{\pm} + \qtr} \pm \half ) r \right) r d\phi  + K_0 \left( (\sqrt{m^2_{\pm} + \qtr} \pm \half ) r \right) d t
\right] \,.
\eea
The coefficients of the inhomogeneous modes on the $S^1$ are
determined by imposing the correct boundary conditions at the
delta function source. At large $r$ all these modes fall off
exponentially, leaving the homogeneous magnetic term $d\phi$.
Note that $dF=0$ everywhere.

\subsection{Membrane source in $\R^{1,6}$}

Write the metric as
\be
ds^2 = d{\bf x}^2_{1,2} + dr^2 + r^2 d\Omega^2_{S^3} \,.
\ee
Place the source membrane along the ${\bf x}$ directions. We want
to solve the following equation for the four form $G$
\be
d \star G + G + 2\pi^2 \d(x) = 0 \,.
\ee
Splitting $G$ into an electric and magnetic part
\be
G = A \, dt \wedge dx \wedge dy \wedge dr + B \, dr \wedge r^3
\text{vol}_{S^3}
\,,
\ee
it is easy to solve the equation in terms of Bessel functions
\bea
A & = & \frac{K_1(r)}{r^2} + \frac{K_0(r)}{2 r} \,, \nonumber \\
B & = & - \frac{K_1(r)}{2r} \,.
\eea
Both of these functions decay exponentially at large $r$. At small
$r$ there is electric flux due to the `Maxwell' term. The phase
$\int_{B^4_r} G$ goes from $0$ as $r \to 0$ to $-2\pi^2$ as $r \to
\infty$. Note that $dG = 0$ everywhere.

\subsection{D string source in $\R^{1,4}$}

Write the metric
\be
ds^2 = d{\bf x}^2_{1,1} + dr^2 + r^2 d\Omega^2_{S^2} \,,
\ee
and place the D string source along the ${\bf x}$ directions. We
need to solve the following coupled equations
\bea
d \star H_3 + \tilde F_3 = 0 \,, \nonumber \\
d \star \tilde F_3 - H_3 + 4 \pi \d(x) = 0 \,.
\eea
It is straightforward to obtain the solution
\bea
\tilde F_3 & = & \frac{(1+r) e^{-r}}{r^2} dt \wedge dx \wedge dr \,, \nonumber \\
H_3 & = & \frac{e^{-r}}{r} dr \wedge r^2 \text{vol}_{S^2} \,.
\eea
Again we find an exponential decay at large $r$. The D string has
sourced $\tilde F_3$ electrically as usual, but has also sourced
$H_3$ magnetically.

\end{document}